# Conceptual Data Modeling: Entity-Relationship Models as Thinging Machines

Sabah Al-Fedaghi
*salfedaghi@yahoo.com*
Computer Engineering Department, Kuwait University, Kuwait

**Summary**
Data modeling is a process of developing a model to design and develop a data system that supports an organization's various business processes. A conceptual data model represents a technology-independent specification of structure of data to be stored within a database. The model aims at providing richer expressiveness and incorporating a set of semantics to (a) support the design, control, and integrity parts of the data stored in data management structures and (b) coordinate viewing of connections and ideas on a database. The described structure of the data is often represented in an entity-relationship (ER) model, which was one of the first data-modeling techniques and is likely to continue to be a popular way of characterizing entity classes, attributes and relationships. This paper is an attempt to examine the basic ER modeling notions to analyze the concepts to which they refer as well as ways to represent them. In such a mission, we apply a new modeling methodology (thinging machine; TM) to ER in terms of its fundamental building constructs, representation entities, relationships and attributes. The goal of this venture is to further the understanding of data models and enrich their semantics. Three specific contributions to modeling in this context are incorporated: (a) using the TM model's five generic actions to inject processing in the ER structure; (b) relating the single ontological element of TM modeling (i.e., a thing/machine or *thimac*) to ER entities and relationships; and (c) proposing a high-level integrated, extended ER model that includes structural and time-oriented notions (e.g., events or behavior).

## 1. Introduction

According to Silvert [1], one of the universal scientific activities is modeling. In scientific contexts, a model is a simplification of a phenomenon that provides insights about the object of study and contributes to its understanding [2]. In engineering, it is now established that models are useful means of understanding and interacting with both products and processes [3]. According to Suppes [4], models are not just representations; they are also tools constructed with an eye toward achieving specific practical purposes

### 1.1 Data models

In software engineering, software development models often represent a networked sequence of activities, objects and transformations utilizing certain notation, syntax, or semantics suitable for computational processing [5]. Models serve in visualizing the design of processes, aiding in idea generation, problem solving and evaluation and in facilitating the interaction and communication [5]. However, in this context, modeling is a quite recent development, and phenomena that can be expressed in models and ways models can be used and interpreted are still not fully understood [6].

In this paper, we take an interest in data models in database systems. According to Elmasri and Navathe [7], data are known facts that can be recorded and that have implicit meaning. Data are sometimes defined as the representation of facts, concepts or instruction in a formal manner that is suitable for understanding and processing. Data can be represented in symbols such as alphabets (A-Z, a-z), digits (0-9) and special characters (+,-.#,$, etc.); for example, 25, "ajit" etc. [7].

We find many data models with the aim of providing increased expressiveness and incorporating a richer set of semantics into a database [8]. Databases require a data model that facilitates the expression of consistency requirements and depicts information semantics. The data model stresses the required information and the ways it should be coordinated as well as illustrates the design, control and integrity parts of the data stored in data management structures, where the role of a conceptual model coordinates the ability to see connections and ideas in a database. [9].

However, according to West [10], data models can be difficult to read unless considerable care is taken in laying them out. Typically, a data model is contrasted with process models where "tension and confusion" occur between those who have a data focus and those who have a process focus [10]. Most people find process models more natural and therefore see data models as "difficult"; however, it is essential that the process and data models are mutually supportive [10]. The object process methodology [11] is based on the paradigm that views processes and objects as equally important in the system model. The thinging machine (TM) model (see [12] and its sources) goes further by viewing projects and processes as two faces of a dual ontological element called a *thimac* (thing/machine). The TM approach enforces finding the core processes that the organization performs as a first requirement in modeling. A unique contribution of TM modeling is identifying generic processes (actions) upon which all other processes can be built. Thus, the mission in this context is reduced to allocating processes in the system structure because their descriptions are already available in terms of these generic actions.

### 1.2 Conceptual data modeling

The conceptual data model represents the overall structure of data that are independent of any database or physical storage structure. According to Tupper [13], conceptual data modeling is one of the most powerful and effective analytical techniques for understanding the information required to support any organization. An example of such modeling is the ER model, which employs basic constructs that include entities, attributes and relationships. According to West [10], if one has not provided the necessary entity types, relationship types, and attributes in a data model, then the data cannot be held. The term *conceptual* in data modeling refers to the things in the business and the relationships among them, rather than data about those things: "So in a conceptual data model, when you see an entity type called car, then you should think about pieces of metal with engines, not records in databases" [10].

One problem intrinsic in data modeling of any enterprise is "the difference between the human's perception of the enterprise and the computer's need to organize the structures in a particular way for efficient storage and performance" [8]. This raises the possibility of distinct levels of modeling that span several planes of abstraction, from conceptual through to logical and physical [14]. This paper focuses on the conceptual/logical levels of data models. A conceptual data model and a logical data model can be very similar, or even the same, for the same subject area, depending on the approach that is taken with each [10].

Conceptual models are technology independent and can be used for discussions with business people, allowing the concepts in the domain to be represented, discussed and agreed upon. The logical model gives a formation to shape the basis of the physical model and elaborates the conceptual model, adding more detail but still typically remaining technology neutral, allowing analysts to discuss and agree on logical structures [9][14]. In general, data models use two types of data modeling techniques: (a) entity relationship (ER) modeling and (b) Unified Modeling Language (UML) [9]. The object-relational model is a blend of the object-oriented model and the relational model, and it fills the gap between the relational model and object-oriented models. However, "The issue with this model is that it can get intricate and hard to deal with" [9]. This paper concentrates on ER modeling.

### 1.3 Entity-relationship model: Review

ER modeling was one of the first data modeling techniques to be developed [15] and it is likely to continue to be a popular way of characterizing entity classes, attributes and relationships [16]. According to Spaccapietra et al. [17], the extended entity-relationship (EER) models are most frequently seen as offering the richest semantic expressiveness in conceptual modeling. An EER "represents the structure or foundation of your database solution. The EER is so important; I can't imagine designing even the smallest project without one" [18]. The ER model has also evolved into one of the most important structural data analysis and design techniques. According to Kashyap [19], the ER approach pre-empts the approach that is based on the assumption of only examining processes, transactions, outputs or data flows of a system. This last approach gives partial information about the environment of the system. Alternatively, the ER approach helps one to arrive at a true or complete picture of the real world for which database is to be built, and it involves the identification and definition of entities of the concerned real world, entity grouping and description, keeping in view the problem area context [19]. An ER diagram is utilized as a pictorial instrument for addressing the ER model, assisting in communicating ideas to a wide range of stakeholders because of its simplicity. In the ER model, platform-specific information and other implementation details, such as procedures, are excluded [9] [14].

In the ER model, an entity and entity type is defined within the context of the organization whose database is supposed to be built. An entity in ER is something that involves information. It is usually identifiable. Each entity has certain characteristics, known as attributes. A grouping of related entities becomes an entity set [20]. Classifying entities into groups and subgroups depends on such factors because the entities can play a role in one organization but not in another [21]. Additionally, the ER model supports a relationship between the entity types themselves.

Yarlagadda and Syed [9] described the ER model as a blueprint of a database that can be actualized as a database. It is a high-level data model that defines data components and their relationships. In many cases, ER modeling is used to produce a semantic data model for a relational database and its requirements. According to Yarlagadda and Syed [9], an ER model can be handily changed over the relational model. The relational model has a modeling methodology independent of the details of the physical implementation; According to Osborne [18], "For example, an *entity* is a person and a collection of people would be a table with each record in that table holding the information describing a unique person. A field in the people table might be hair color, height or anything describing a person for which there is one choice."

According to Yarlagadda and Syed [9], an industry standard for building up an ER model does not exist, and developers may utilize documentation that is not perceived by different designers. The model does not offer a sufficiently rich conceptual model for problems that do not map onto tables in a straightforward fashion [8]. According to West [10], ER diagrams have sometimes been found "confusing" and limited in what they can express, as follows.

- If one entity type happens to be an instance of another, then there is no way to say this in an ER diagram. The problem with this is that the relationships are really classes of a relationship. For example, in *Each pump is a member of exactly one equipment type*, it is not the entity type pump that is an instance of equipment type, but an instance of pump that is a member of an instance of equipment type.
- One cannot show a subtype relationship between an entity type and an instance of any entity type in the data model.
- Entity type definitions are often ambiguous or unclear [21].

Additionally, according to West [10], real-world relationships do not automatically align with the lines in the ER models, and hence it would be confusing to use the word "relationship" for both of them. In the ER technique, relationship types (lines) are second-class objects, in that they are dependent on that to which they relate. According to Osborne [18], "Relationships are the easiest and hardest thing to understand. They are based on a simple mathematical equation but so many people, including myself, struggle with them. What helped me understand them better [is that] the relationships are just questions you are asking your data. The question is expressed in terms of a mathematical equation but is still just a simple question like; show me the records that equal this value. It's really that simple."

### 1.4 Outlines

The ER model has undergone a variety of changes and extensions over the years [22]. However, the focus in this paper is on the basic ER model as presented in Chen [23]. The paper applies a new modeling methodology, TM, to the ER model in terms of its fundamental building constructs, representation methodologies for the objects and the methods to express the semantics of the application environment. The goal of this venture is to suggest enriching the semantics of the ER model by reducing all operations to five generic actions and providing a basis for the notion of system behavior at a level above the static description level.

Accordingly, the following is an outline of this paper:
- In Section 2, we review the TM model, which forms the theoretical foundation of work in the paper.
- Section 3 provides illustrations of TM representations: sets, subsets, individuals and relationships. The section can be thought as an introduction to applying TM to ER notions.
- In Section 4, we show how to build a TM model that includes events and behavior. We concentrate on basic operations (e.g., inserting a tuple that preserves the integrity constraints in the database schema).
- In Sections 5 and 6, we further pursue our objective of translating ER diagrams into the TM model. We introduce a case study that represents a more complicated ER diagram.
- In Section 7, we express functional dependency in a TM model.

### 2. Thinging Machine Modeling

Schematic models are abstractions of reality that are developed to understand things and processes. The TM model is a conceptualization of how things/processes can be merged into a complex of interrelated thimacs (i.e., things that are simultaneously machines). All things (thimacs) are created, processed, and transported (acted on), and all machines (thimacs) create, process, and transport other things (thimacs; see Fig. 1). Things "live" or "pass through" other machines. Machines house other things and provide roads for their flow. The unity of thing and machine forms a thimac. In such a blend, a single thimac is a fusion of two manifestations. The thing flows within machines while it is a machine for other flowing things. A complete machine is shown in Fig. 2. The machine in Fig. 2 is more complete than the known input-process-output model. For example, suppose that we study the productivity of a particular organizational unit. It is not sufficient to examine only the output; we also have to examine what is being created. What is created may or may not be output.

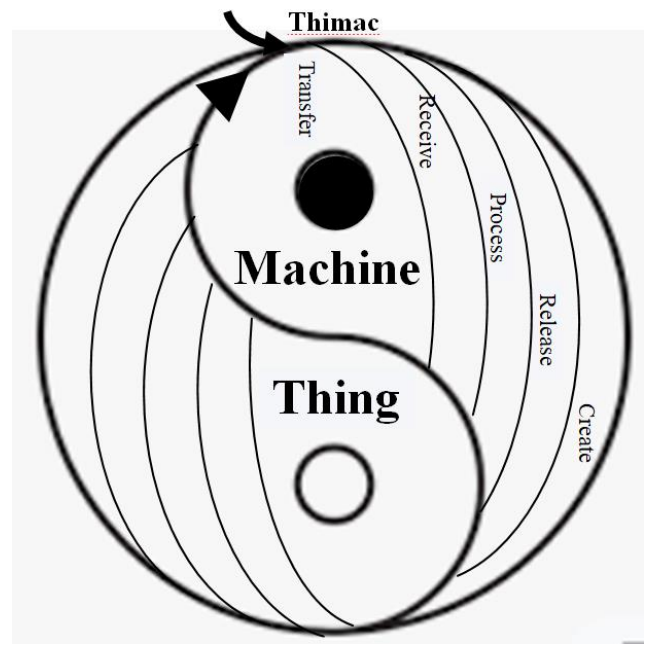

Fig. 1 A yin-yang symbol is a circle divided by an S-shaped line into two segments representing a thing and machine, each being a version of the other. A thing flows into a machine and a machine becomes a thing.

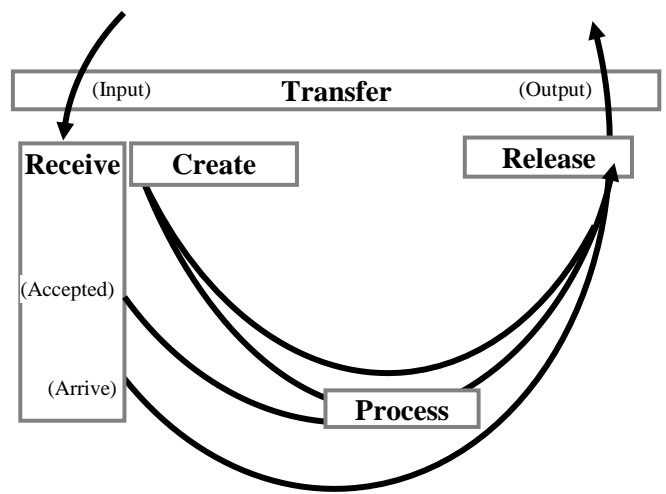

Fig. 2. Flow of things in a thinging-machine model

As shown in Fig. 2, a TM machine can be viewed as a coordinated system of flow (a change in the action position). The flow is not a type of link (e.g., a class of relationship in ER); it is rather a transformation from one state to another. Fig. 2 can be described in terms of the following generic (has no more primitive action) actions:

**Arrive**: A thing moves to a machine.
**Accept**: A thing enters the machine. For simplification, we assume that all arriving things are accepted; hence, we can combine the arrive and accept stages into one stage: the **receive** stage.
**Release**: A thing is ready for transfer outside the machine.
**Process**: A thing is changed, but no new thing results.
**Create**: A new thing is born (being found/manifested) in the machine. Things come into being in the model by "being found." Creation in metaphysics involves bringing the entities from the state of nonbeing into existence. The TM model limits this creation to the appearance in the model. *Create x* in a model means "there is" *x*. After the instance of creation, the entity may move to be processed or released, or it may stay in the creation state.
**Transfer**: A thing is input into or output from a machine.

Additionally, the TM model includes the mechanism of triggering (denoted by a dashed arrow in this study's figures), which initiates a flow from one machine to another. Multiple machines can interact with each other through the movement of things or through triggering. Triggering is a transformation from one series of movements to another.

To conceptualize a thimac as a thing presents no indication as to the content of the thing, whereas to conceptualize a thimac as a machine forces a definite structure of actions with flow of other things. The thimac is a device to *district* a segment (this term is taken from [24]) from the other segments in the universe. The totality of the universe is also a thimac. In TM, a set and relationships are also thimacs. For notational convenience, they can be drawn differently. Thus, a binary ER relationship, say (customer, product) is a thimac that includes two subthimacs, customer and product, with their machines, as will be illustrated in the next section.

An important distinction in TM is between static thimacs and events. We would expect that the static description as an organizational (structural)/formational/topographical level that does not specify the instances or events. In the static form (e.g., TM diagram/subdiagram), everything is there; nothing corresponds to time (past, present or future), and nothing corresponds to, say, the principle of no contradiction. However, what is "there" is loaded with potentiality that can be exemplified by actuality. For example, *an order of a product, its processing and its response (negative or positive)* are present in the picture side by side. An Event is a static thimac that has a time "breath" (subthimac) that infuses dynamism in the thimac, just as creating a wave is the static ocean. Dynamism is a regulating mechanism of the static form to align it with reality through such machinery as igniting and chronologizing actions, logicalizing, and executing/controlling processes. For example, *Create x* as a static description is analogous to *create Pinocchio*, but the event gives Pinocchio the ability to flow in the model. Dynamism involves the development of actuality and the realization of static form through time.

Accordingly, a thing with a time subthimac is considered an instance (individual). Individuals are things that exist in space *and* time. Examples (taken from Milton and Kazmierczak [25]) of individuals are as follow:

- An accountant named Freda qualified with the *time* of being an accountant and an individual named Freda.
- The annual financial statement for Ericsson is an individual if it is fixed (created and continued) with a specific *time*.
- Orly International Airport is an individual from the *time* of its construction.

## 3. Illustration of Thinging Machine Representations: Sets, Subsets, Individuals and Relationships

In this section, we begin to explore the expression of ER notions using TM. The examples in this exploration also introduce TM modeling of ER examples in the coming sections. We discuss two issues in this section:

- TM, as an engineering diagram, includes all needed specifications for actual realization. If simplification is needed, then TM can be reduced to the ER description.
- The TM model distinguishes between the static and the dynamic levels by drawing time events over the original static diagram. The ER counts time as an attribute of an entity.

**Example**: Consider the example described in Green [26] as an instance of relationship of association among two or more entities: *"Customer 'Smith' orders product 'PC42.'"* In TM modeling, we find that the given relationship is replaced with static and dynamic machines as shown in Figs. 3 and 4, respectively.

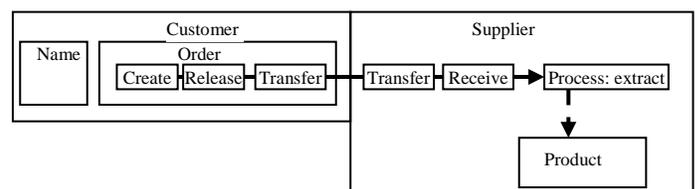

Fig. 3 The static TM model of relationship between Customer and Product

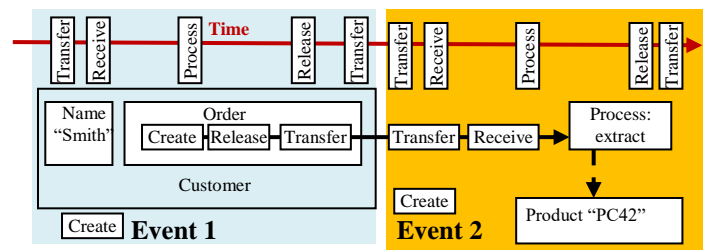

Fig. 4 The relationship 'Smith' orders product 'PC42' as two events in TM

In Green [26], the time in *"Customer 'Smith' ordered product 'PC42' on January 11, 2005* is just an *attribute*, but in TM modeling, the time moves the data *"Customer 'Smith' ordered product 'PC42'* to the level of realization as an event. Conceptually, this is an important difference between the ER model and the TM. From the TM point of view, the ER representation is a static representation that is incapable of modeling dynamism. TM modeling has two levels of abstraction: static and time. *January 11, 2005* is a thimac located at a higher level of abstraction, whereas *'Smith' ordered product 'PC42'* is a region (of the static TM diagram) that has an association with the time *January 11, 2005*. Thus, *January 11, 2005* "lifts" the thimac *'Smith' ordered product 'PC42'* from the static level to the level of dynamism (events).

**Example**: Karukonda [27] provides a classic example of ER modeling shown in Fig. 5 that represents *the red colored ball broke the glass window*. Karukonda commented that the example indicates the semantic primitives provided by the ER approach are more fine-grained than the semantic primitives provided by a semantic net are. The ER approach maintains a distinction between entities, relationships, roles, attributes and values. Fig. 6 shows the corresponding TM model. On the left of the figure, we find the ball and its attribute color. The ball moves to the window, specifically, to the glass part of the window "causing" the break. Comparing the two representations, we note that the TM representation includes structure and actions. If we remove the five TM actions, we obtain Fig. 7, which can easily be converted to the ER diagram where "break" forms the relationship. We can conclude that the relationship "break" in the ER graph is replaced by actions and the state "break" (or being broken) in the TM model.

We can see that the relationship in this example stands for actions in TM. If we look at the corresponding language expression, ER representation stands for *the red ball breaks the glass of the window*. The TM representation stands *for the red ball flies and hit the widow causing the glass to break*. We leave the issue of which representation has, in Karukonda's [27] words, "more fine-grained semantic primitives."

It is generally claimed that the ER diagram represents the conceptual structure of a problem domain being modeled (e.g., [28]). As this example shows, the actions of such a model are sometimes hidden under the relationship. As we will see later in this paper, what is called *relationship* in this example contains the roots of behavior of things (what the thing does) where events are the product of the encounter between a series of actions and time. Thus, behavior is an intrinsic integration of actions and time. Fig. 8 shows the generic actions in *the red colored ball broke the glass window*. Fig. 9 shows the chronology of these generic events.

In the ER model, behavior is defined through another mechanism (e.g., Petri nets), such as a set of states and transactions that bring about state transitions. Nevertheless, the TM version of the ER diagram tells us there are embedded actions hidden by the ER relationship. Actions *mean* change and movement; hence, these in time lead to events. TM representation exposes what is hidden in the ER notation and similar other models, such as UML activity diagrams, and those representational methods are too weak to model behavior.

It is interesting to explore this notion of event using TM representation. Suppose the event $E_j$ happens after $E_i$. This succession of events can be modeled in TM. In TM modeling, an event is a *fuzzy* concept. According to this TM view, all thimacs are placed with time as an order of succession, as well as in space (or staticity: subdiagrams of the TM static model) as an order of situations. Such a statement imitates Newton famous statement, "All things are placed in time as to order of succession; and in space as to order of situation" [29].

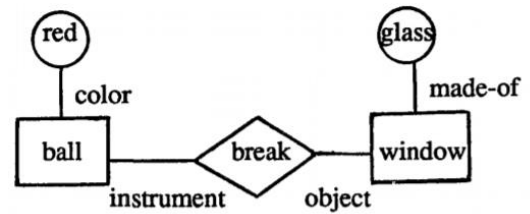

Fig. 5 The ER model of *the red colored ball broke the glass window*

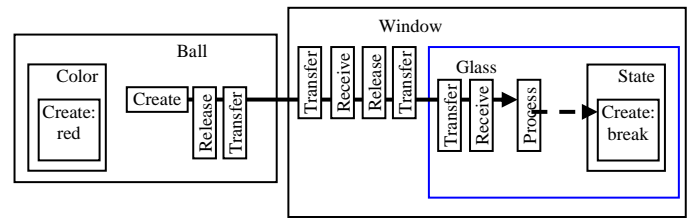

Fig. 6 The static TM model of the red colored ball broke the glass window

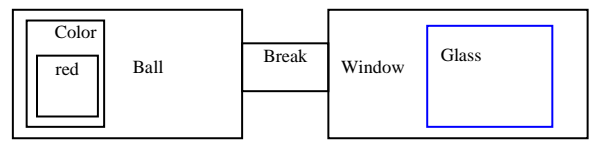

Fig. 7 Removing the TM actions

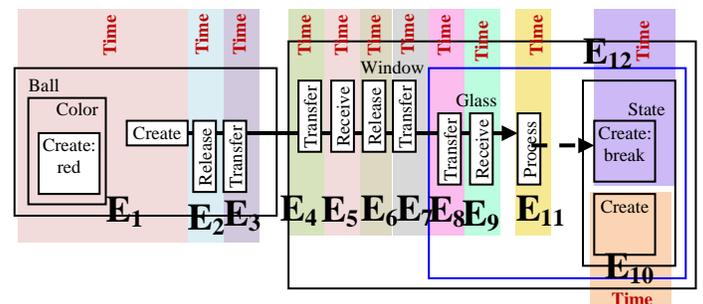

Fig. 8 Events is an intrinsic integration of actions and time

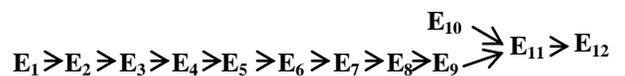

Fig. 9 Behavior of *the red colored ball broke the glass window*

In this sense, we understand that the staticity (i.e., the TM subdiagram) *exists* (i.e., *create* in the diagram) and is resting in this existing time (create) but happening *happens* in the flowing time (transfer, receive, process, release and transfer; see Fig. 10).

**Example**: Consider the activity diagram shown in Fig. 11. The semantics of the arrow in this diagram raise some question. We can distinguish between logical sequence and time sequence. Since there is no mention of time, we assume there is a logical sequence (e.g., first do *Press Power*, which is followed by *computer does the posting*, then *starts OS*, etc.). The arrows also imply a time sequence, say *Press Power* at time t1 the *Post happens* at t2, t1 < t2, etc. Suppose the event of t1 happened in the year 1980 and the event of t2 happened in 2020. The diagram is still valid. Hence, *practically*, the diagram does not imply a time sequence and the arrows denote only logical sequence as the number sequence 1, 2, 3, 4 implies an order. Contrary to the common claims, we can conclude that the activity diagram has nothing do with behavior since behavior involves change (flow or triggering), time (thimac) and events (staticity plus time). The same conclusion is applied to other modeling diagrams, including ER and even state diagrams. On the other hand, in TM modeling, time, which is a thing, does not allow empty or unoccupied instants.

## 4. Sets and Subsets in ER and TM Modeling

The issue discussed in this section relates to class-based modeling that deals with such matters as classes, subclasses and elements in these classes as well as whether a constraint holds among a given set of classes.

Consider the ER diagram in Fig. 11 that represents the relationship between *customers* and *addresses*. Fig. 12 shows the *customers* part in the corresponding TM model, where there is an explicit distinction between the set (entity type) *customers* and an individual *customer*. Both customers and custom in Fig. 13 have *being* (possibility) [29] in the system. Instances or individuals are realized (have existence; i.e., actuality) when thimacs have time subthimacs as shown in Fig. 14.

That is, an instance of *customers* is created at (and during) a given time. If the time subdiagram is not included, then *customers* is just a static description. The issue here is related to the difference between extension and intension in logic.

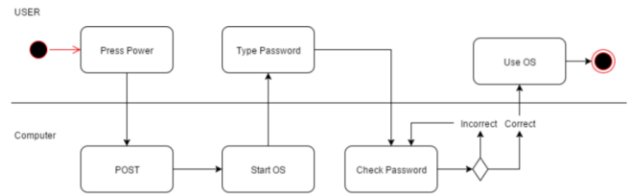

Fig. 11 An activity diagram (from Ruppel [33])

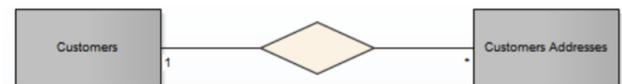

Fig. 12 The relationship between Customers and Customers Addresses (adopted from Sparx [14])

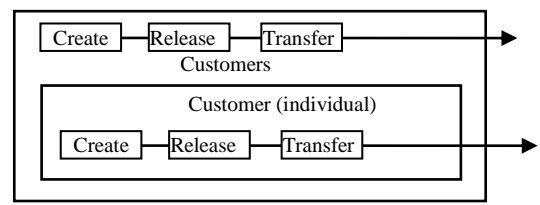

Fig. 13. TM representation of customers

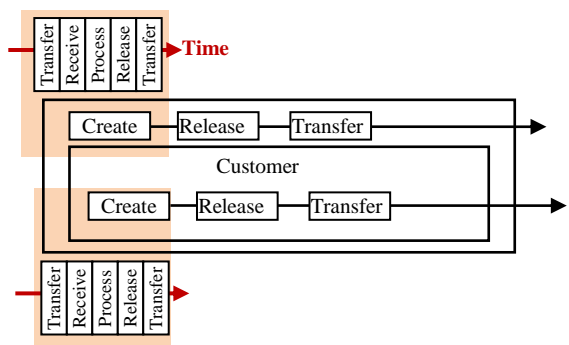

Fig. 14. Instances are realized in time

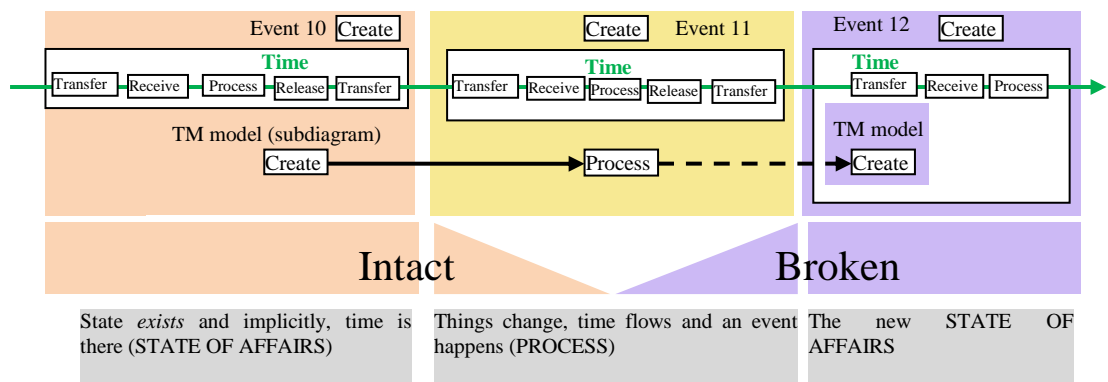

Fig. 10 "What happens in time, and what exists in time; but these two ways of being in time are different" [29].

In TM modeling, all notions should be represented explicitly. If there is a reason to erase the difference between any two notions (e.g., set and subset, for the sake of simplification), then the TM diagram can be simplified starting from the original complete diagram. The idea is that any engineering diagram should be a *complete* diagram ready for realization (e.g., programming). For example, any engineering system can be represented by block diagrams, which are interconnected blocks that each describes a portion of the system. Block diagrams usually adhere to a mathematical system of rules (e.g., in control engineering). A block diagram can be simplified using various methods. For example, block diagrams can be simplified using a signal flow graph [30], which uses the rules of block diagram algebra. Nevertheless, simplification should be based on the original diagram.

In Fig. 13, a set called *customers* includes multiple members where each member is called *customer*. For notational clarity, we can draw double line edges or boxes to denote multiplicity of individuals. For simplicity's sake, we can use a single line box under the assumption that the name of the thimac reflects the distinction between a set and members (e.g., *customers* vs. *customer*). Additionally, the action *create* indicates the appearance of the thimac in the global system. In TM representation, what "exists" is that which appears (is captured) in the TM diagram; or in Gruber's [31] words, "That which can be represented." As will be practiced in this paper, *create* may be omitted under the assumption that the presence of a box with a name is sufficient to indicate the presence (being) of the thing (type) in the system.

The relationship 1-n is also a thimac as shown in Fig. 15. It is drawn as a dotted box for emphasis. It can be created and transferred. If we want to simplify the TM model to extract an ER-like description, then we can eliminate the distinction between a set and a member and then represent the relationship by a diamond shape.

## 5. Integrating Operations in Structure

In this section, we show how to build a TM model that includes events and behavior. We demonstrate that we can build a set of operations using the structural framework of the example developed in the previous section. This is in line with the ontological thesis of the TM that unifies things and machines, structure and processes. Typically, in data models, the operations are defined as algebraic operations. The user specifies the data of interest, and new relations (tables) are formed by applying relational operators. In this section, we concentrate on basic operations such as insert, delete and update that preserve the integrity constraints in the database schema. We focus on such referential integrity specified between two tables and used to maintain the consistency among tuples in the tables. Constraints typically arise from the relationships among the entities in the relational schemas. Note that the main aim of specifying the operations is demonstrating one way of implementing it without regard to such considerations as efficiency or optimum algorithms.

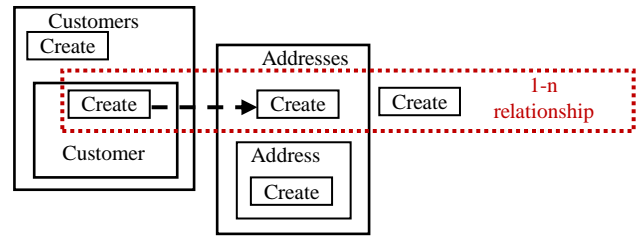

Fig. 15. The relationship is also a thimac

We leave such issue to the design and implementation stages of development. We can say that the aim at the conceptual stage to describe what we want not how to do this in a good way.

### 5.1 Insertion Operation: Static Model

Using the customers/addresses example discussed previously, we assume, without loss of generality, a user request includes two elements of data: the *customer ID* and the customer's *address* to be inserted (see Fig. 16, pink number 1). In the figure,

- The request flows to the system (2) where it is processed (3) to extract the customer ID (4).
- On the other hand, the system has the *customers* file (teal color 5). The customer box does not include the relationship 1-n that was discussed previously because such a constraint is not needed in the operation if inserting an address. In a later example, when we discuss the update operation, the TM representation of the 1-n relationship will be included.
- The file *customers* is processed (6) to extract the record of a single customer and send it to check whether it is the record of the given ID (7, 8 and 9).
- If the record extracted from the file *customers* is not a record of the customer with the given ID (10), then this triggers the extraction of the next record in the *customers* file and the process of comparison is repeated.
- If the record extracted from the file *customers* is the record of the customer with the given ID (11), then the given data (ID and address) are processed (12) to extract the address (13) and rebuild the retrieved record from *customers* (14) by processing the new address and the record (15) to be included in the *customers* file instead of the old record, thus, creating a new *customers* file (16).
- If no record is found that matches the given ID (EOF – 17), then a new record is constructed (18 and 19). Hence, a process (20) inserts the new record in the *customers* file, thus creating a new file (21).

### 5.2 The Events Model

In preparation to develop the behavioral model in TM representation, we identify the events over the static model of Fig. 16. Fig. 17 shows the following events where, for simplification sake, time is not included.

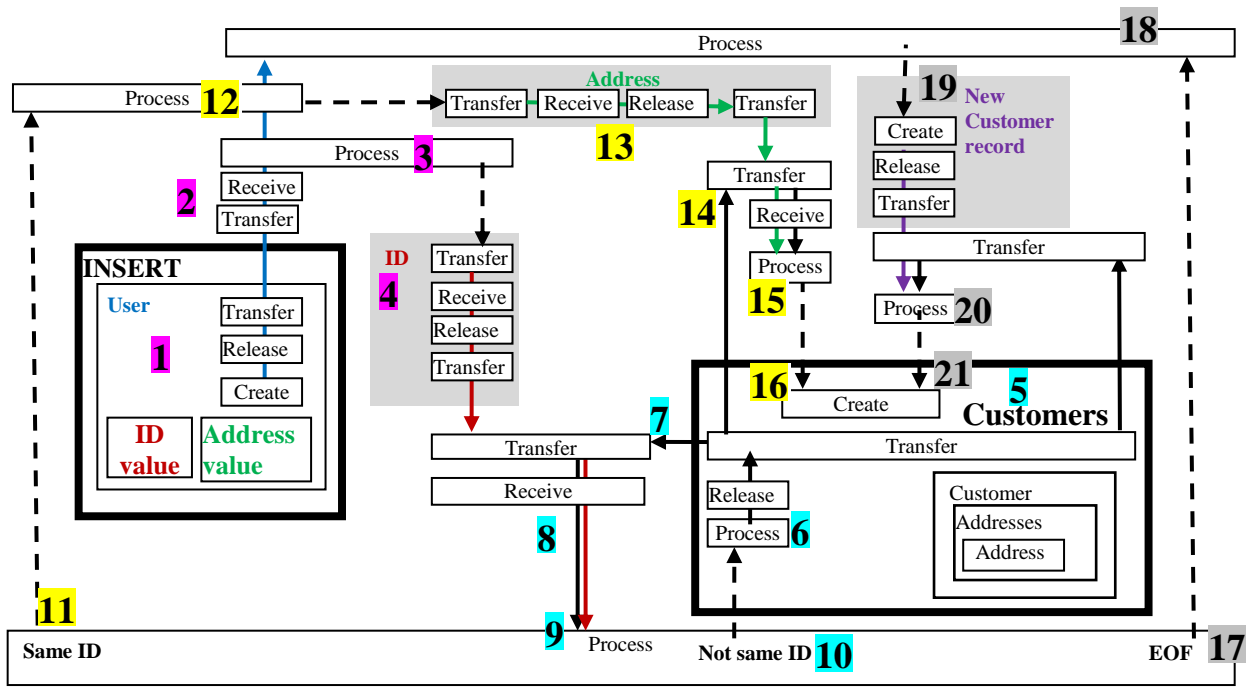

Fig. 16 The TM model of the insert operation

E1: A request to insert a new address for a customer has arrived.
E2: The request is processed to extract the ID. Note that extracting the ID indicated the arrival of the ID (transfer-receive), which was embedded in the request.

E3: The customer file is processed to retrieve a record.
E4: The input ID and the retrieved record from customers are processed to determine whether the record is of the input ID.
E5: The input ID is not the same as the retrieved record.

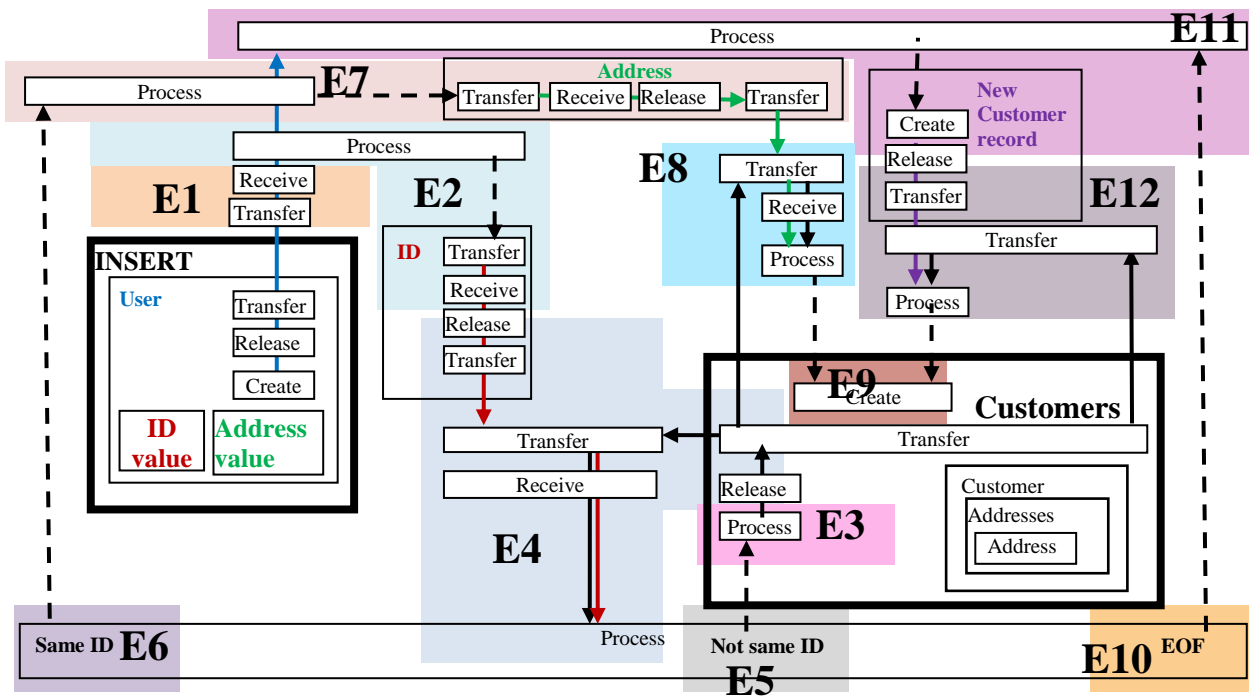

Fig. 17 The TM model of the Insert operation in the customers/addresses example

E6: The input ID is the same as the ID in the retrieved record.
E7: The input request is processed to retrieve the address.
E8: The address and the customer file are processed to add the address to the found record.
E9: A new customer file is created with the updated record.
E10: The ID does not correspond to any record.
E11: The input request is processed to create a new record.
E12: The newly created record and the customer file are processed.
Fig. 18 shows the behavioral model of the customer/address example.

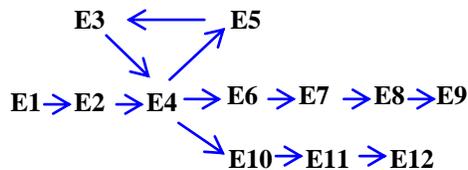

Fig. 18 The behavioral model of *insert an address*.

## 6. From ER to TM

In this section, we further pursue our objective of translating ER diagrams into the TM model. We introduce a case study that represents a more complicated ER diagram taken from Elmasri and Navathe [7] as shown in Fig. 17. The figure involves a simplified sample COMPANY database. The employee number, Ssn, is assumed to be the key for EMPLOYEE in Fig. 17.

### 6.1 Representing roles

According to Elmasri and Navathe [7], the ER model in Fig. 17 includes the following roles.
- There is the WORKS FOR relationship n-1 between EMPLOYEE and DEPARTMENT.
- PROJECT has a 1-n relationship with DEPARTMENT
- EMPLOYEE can work on PROJECT.
- A single employee and single project belong to a single department.

In TM, as discussed previously, it is necessary to distinguish between sets and subsets for reasons related to such relationships as n-1, n-1 and m-n that involve subsets of the sets. Thus (see Fig. 18),
- The set EMPLOYEES can also be represented as a set of subsets SUBEMPLOYEES (see subsection 4.2).
- The set DEPARTMENTS can also be represented as a set of subsets SUBDEPARTMENS, but in this example, it is not necessary in the roles above.
- The set PROJECTS can be represented as SUBPROJECTS.

Fig. 18 shows initial TM specifications of the Elmasri and Navathe [7] simplified ER model. Note that the roles are drawn in color as thimacs. For example SUBEMPLOYEES is a subset of EMPLOYEES. A SUBEMPLOYEE participates in the relationship called n-1 with an individual department. In contrasting Fig. 18 with the ER diagram of Fig. 17, we see that TM explicitly brings a relationship that involves subsets. Note that in Fig. 18 relationships are modeled as thimacs. Thus, we have five additional *top* thimacs (in colors), i.e., top refers to thimacs that are not the subthimacs of higher thimacs. Each of these thimacs represents an ER relationship. We will next describe how the relationship between employee and department is implemented in TM. Fig. 19 specifies this thimac as one individual employee belonging to one individual department. How do we preserve this relationship between any employee and his/her department? First, we consider the operation of inserting an employee in a department.

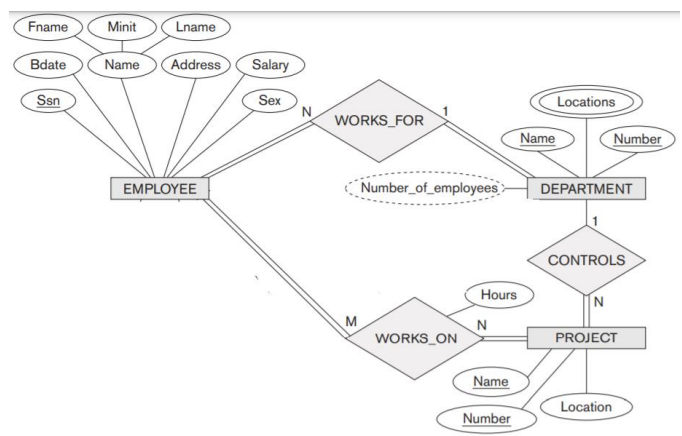

Fig. 17 An ER schema diagram for the COMPANY database (Incomplete from [7]).

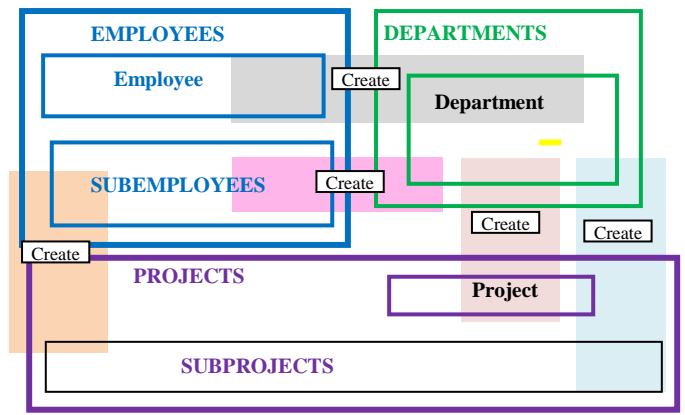

Fig. 18 A TM initial diagram that specifies the relationships

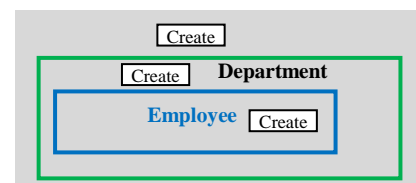

Fig. 19 The role: one individual is in one individual department.

## 6.2 Operations

To implement such an operation as inserting a record of the new employee in a given department, e.g., INSERT (employee NO. [SSN], department ID), we first need to check the following as shown in Fig. 20. Note that for illustrative purposes, the EMPLOYEES box is drawn with double lines to indicate that it is a set.

1. For finding whether a record of the employee is already in the database, we assume that this can be accomplished by comparing the SSN with every record in EMPLOYEES, the set of all records in the database. Note that at this conceptual level, there is no consideration for efficiency. Hence in Fig. 20, after receiving the insertion request (pink number 1), the SSN is extracted (2) Additionally, the database EMPLOYEES (3) is processed (4) to retrieve a single record (5) that is processed (6) to extract its SSN. The employees' numbers coming from the insertion request (2) and the one extracted from the database record (6) are compared (7). Accordingly,
   - If they are equal, this indicates an error because the employee already has a record in the database (8).
   - If the two employee numbers are not equal, then the database is processed to retrieve the next record in the database (9).
   - If at the end of the file (EOF), then this indicates that the new employee is currently in the database; thus, the insertion process continues (10).
2. A new employee record is constructed (11).
3. Using the department number DeptN that is extracted from the insertion request, the file of EMPLOYEES in DeptN is identified (12, 13 and 14).
4. The file of DeptN EMPLOYEES and the new record are processed to insert the new record into the file (15, 16 and 17).
5. A new file for EMPLOYEES of DeptN is created and replaces the current file (18).

Fig. 21 shows the corresponding events diagram divided according to the following events.
$E_1$: The insertion request is received and processed.
$E_2$: The SSN is extracted.
$E_3$: EMPLOYEES is processed.
$E_4$: A single record is retrieved.

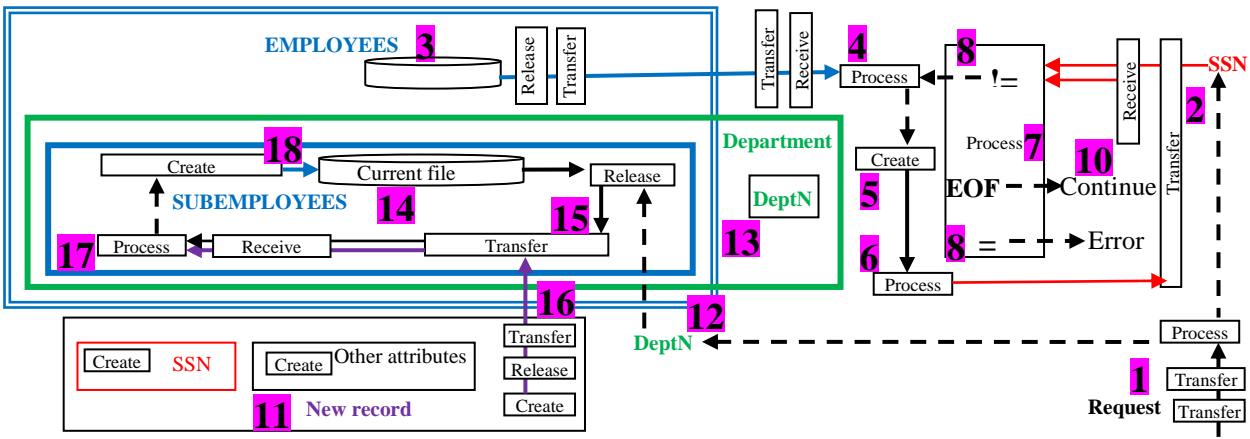

Fig. 20 The TM static model

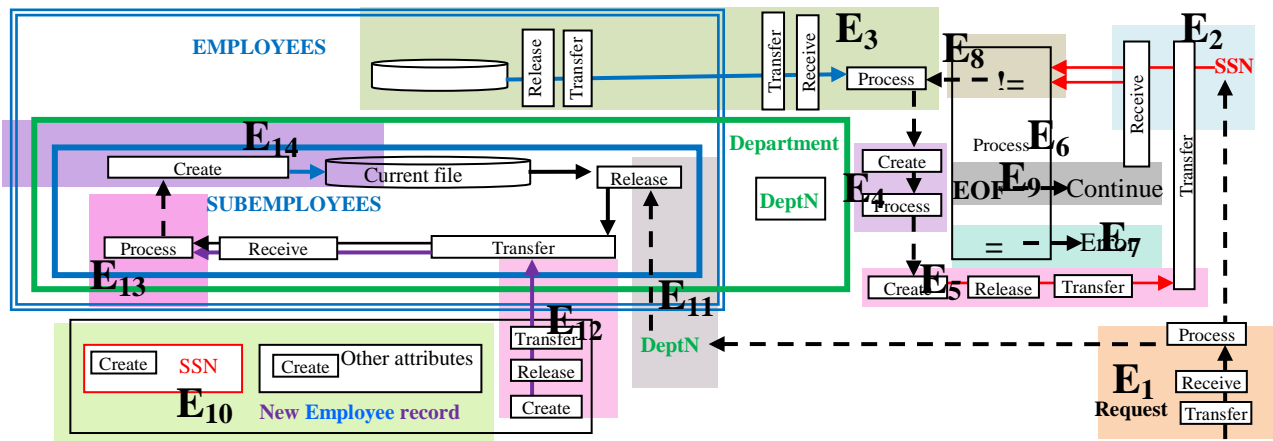

Fig. 21 The TM events model

$E_5$: The Ssn in the record is extracted.
$E_6$: The two employee numbers from the insertion request and the retried record are compared.
$E_7$: The two numbers are equal, hence Error.
$E_8$: The two numbers are not equal; hence, another record is retrieved.
$E_9$: EOF; hence, continue.
$E_{10}$: A new employee record is constructed.
$E_{11}$: the department number is extracted and the EMPLOYEES of the department are released.
$E_{12}$: The constructed new employee record flows to be processed.
$E_{13}$: The new employee record is inserted in the EMPLOYEES of the department.
$E_{14}$: New EMPLOYEES of the department are created and replace the current one.
Fig. 21 shows the behavior model of the request for employee insertion.

It is interesting to analyze again the relationship in which an employee belongs to a single department. In Fig. 18, this relationship is drawn as a machine that includes employee (individual) and department (individual). However, in Fig. 20, the relationship between employee (individual) and department (individual) is constructed into two steps:
1. check that the employee does not exist in the database, and if this is true, then,
2. insert the employee into the subset of employees in the department.

The first step guarantees that no other department includes the new employee. The second step asserts the presence of the new employee in his/her department. Accordingly, the relationship between the employee (individual) and department (individual) has two requirements: the employee belongs to a department, and he/she does not belong to any other department: Employee x belongs to department y and only y. Hence, one solution to the *specification* of the involved relationship is shown in Fig. 22, which is a subdiagram of Fig. 18 with the subthimac (attribute) uniqueness added to the department. The point of such a discussion is to demonstrate that the TM model presents a more refined specification of some ER notions.

## 7. Modeling functional dependency

The ER modeling is used to build the conceptual model of the relational database. In this context, dependencies—e.g., functional dependencies (FD)—are used to 'normalize' these relations (tables). ER diagrams and normalization are, generally, discussed as two independent concepts with no link between them [32]. According to Dhabe et al. [32], normalization should be viewed as a process of refining ER diagrams as these two concepts are so closely related to each other. Existing ER-diagrams cannot accommodate FD information; this makes it compulsory to enter FD information at some later time with user interaction.

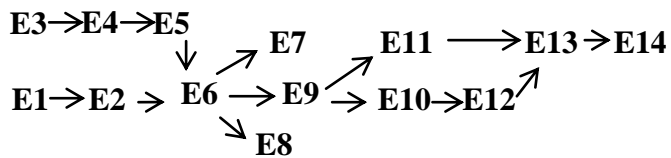

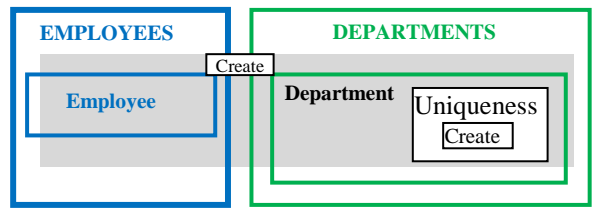

Fig. 22 The TM behavior model

Fig. 23 In the relationships, an employee belongs to a single department, the department should qualified as unique

Any modifications made to the attributes of entities in existing ER diagrams may lead to inconsistent descriptions of FD information and attribute descriptions of entities [32]. Additionally, an automation of normalization would be achieved if a single diagram (e.g., ER plus FDs) is used [32]. Apparently, this problem is the result of limiting the conceptual model to structural aspects of the system. FDs are constraints on the behavior of the system that can be expressed in terms of behavior.

**Example**: Assume the data model of the relation R (Employee ID, Start_Date, Employee Name, Address, Telephone, DOP, Age) shown in Fig. 24 subject to the FDs [32],
1. Employee_ID, Employee_Name → Start_Date
2. Employee_Name → Address, Telephone-Number, DOB
3. Employee_ID → Dependent_Name

Fig. 25 shows a simplified TM representation where actions are deleted. Such a figure includes the attributes and functional dependencies. Each FD is expressed as a two-tuple constraint where the same two values of a subset of attributes of any two tuples trigger the same values as another subset of attributes.

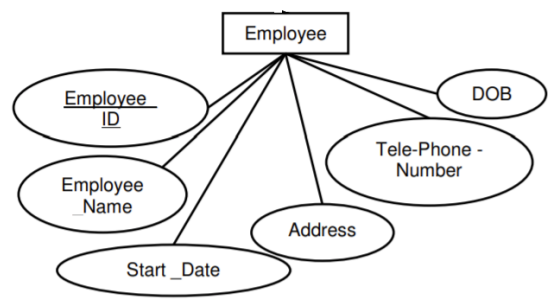

Fig. 24 The entity diagram of Employee (incomplete form [32])

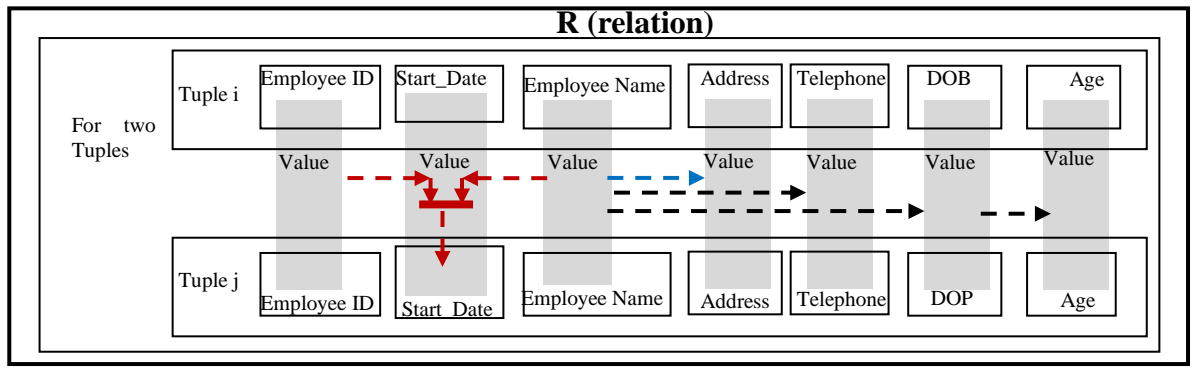

Fig. 25 The relation R (Employee ID, Start_Date, Employee Name, Address, Telephone, DOP, Age) where any two tuples have functional dependencies as shown.

As mentioned in previous sections, all types of roles and constraints are viewed in TM as thimac. As an example, Fig. 26 shows the definition of Employee_ID, Employee_Name → Start_Date.

Consider the operation of *updating* by inserting a new value of address given the *employee Name* that preserves the FD Employee Name → Address (Blue dashed arrow in Fig. 25). This involves the following operations:
- Finding tuples with **Employee name** value that are equal to the given value of the Employee Name (call the set of these tuples R2). R2 tuples have a single **Address** value since R already obeys the FD Employee Name → Address.
- Replacing the values of the **Address** in R2 by the given new value of the **Address**.

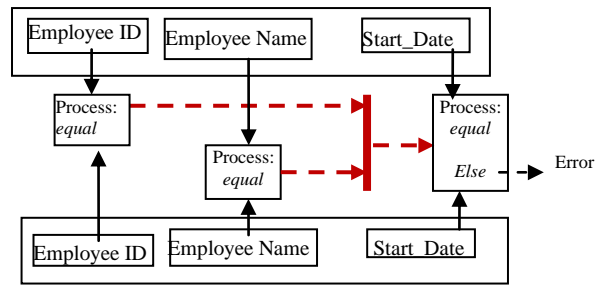

Fig. 26 The machine of Employee_ID, Employee_Name → Start_Date

Fig. 27 shows the TM representation of such an updating operation as follows.

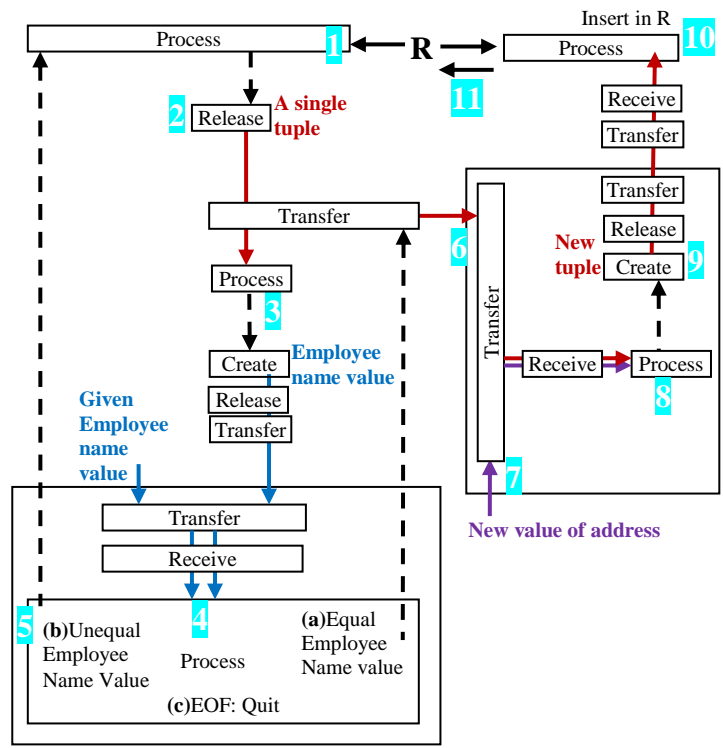

Fig. 27 The TM representation of enforcing Employee Name → Address when performing the operation that updates **Address.**

- First R (the relation/table) is processed (number 1 in the figure) to extract a tuple (2) that is, in turn, processed to extract the value of the Employee number in the tuple (3).
- The extracted value of the **employee ID** is processed (compared - 4) with the given **employee ID value**.
- Accordingly,
  - If the two values are equal (5), the next tuple in R is released (2).
  - If the two values are equal then the tuple and the input new value of the **Address** are processed (6, 7, 8) to construct a new tuple (9), that is inserted in R in place of the old tuple (10 and 11).

Fig. 28 shows the events diagram of this segment of enforcing the FD **Employee Name** → Address and Fig. 29 show the corresponding behavioral model.

The results in this example show the viability of the TM language to express operations while preserving constraints. This provides a tool to analyze the semantics of the ER model that involves FDs and furthers the general understanding of data models and enriches their semantics. The main specific contributions of modeling in this context are incorporating the TM five generic processes to inject processing in the ER structure.

## 8. Conclusion

In this paper, we have examined the basic ER modeling notions to analyze the concepts they refer to, and we have represented them. We apply a new modeling methodology (thinging machines, TM) to ER in terms of its fundamental building constructs, representation entities, relationships and attributes. The goal of this venture is the further understanding of data models and enrichment of their semantics. The preliminary results indicate that TM introduces a new language to express ER notions at a finer level of detail. Specifically, infusing the ER diagram with TM actions presents a combined conceptual picture of structure and actions. Additionally, the clear TM separation of the static and dynamic levels provides a comprehensive framework to develop behavior.

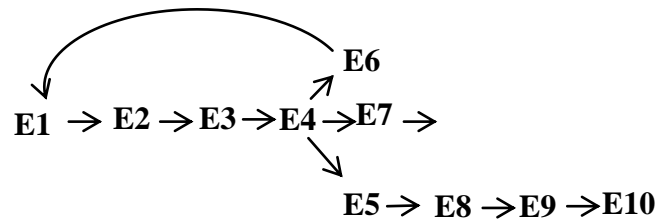

Fig. 29 The behavioral model that realizes the FD.

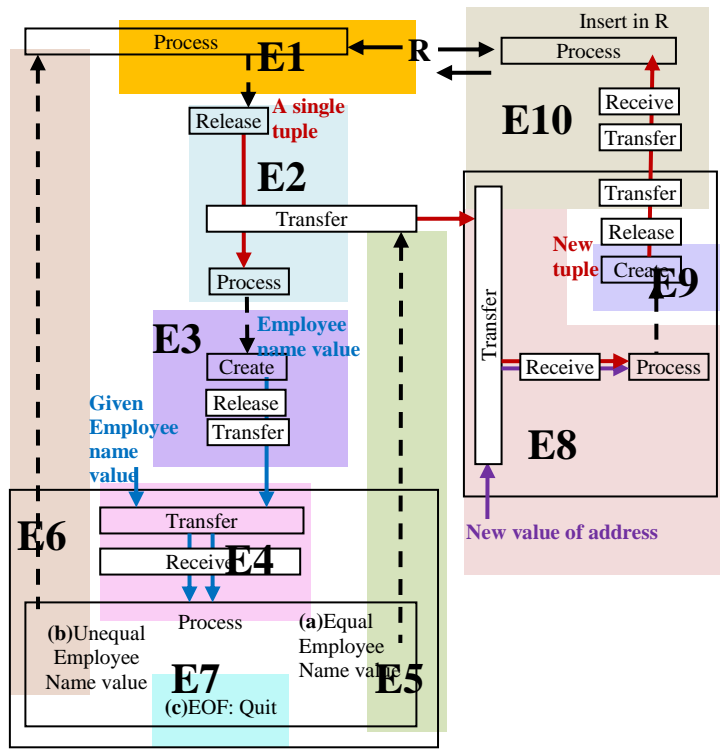

Fig. 28 The events model of enforcing Employee Name → Address when performing the operation that updates **Address.**

As the ER model includes a large body of work and research materials, especially in the area of using the ER model as a conceptual foundation for the relational database model, future work on applying TM to the relational database model would provide an extension to this paper, and the results in this study seem to point to the benefit of further investigation in this direction.

**References**

[1] W. Silvert, "Modelling as a discipline," International Journal of General Systems, vol. 30, no. 3, pp. 261-282. April 2000. DOI: 10.1080/03081070108960709

[2] J. Beal, "From the mechanistic modeling of signaling pathways in cancer to the interpretation of models and their contributions: clinical applications and statistical evaluation," Ph.D. Thesis, University of Paris, September 2020. file:///C:/Users/salfe/Downloads/InstitutCurie_JonasBEAL_2020%20(1).pdf

[3] R. Frigg and S. Hartmann, "Models in science," in The Stanford Encyclopedia of Philosophy (Spring 2020 Edition), ed. E. N. Zalta <https://plato.stanford.edu/archives/spr2020/entries/models-science/>.

[4] P. Suppes, "Models of data," in Logic, Methodology and Philosophy of Science: Proc. of the 1960 International Conference, eds. E. Nagel, P. Suppes, A. Tarski, & 7 more, Stanford University Press, January 1, 1962.

[5] J. J. Marciniak, ed., Encyclopedia of Software Engineering, 2nd Edition, John Wiley and Sons, Inc, New York, Dec. 2001.

[6] C. M. Eckert and M.K. Stacey, "What is a process model? Reflections on the epistemology of design process models," in: Modelling and Management of Engineering Processes, eds. C. M. Eckert and M.K. Stacey, Modelling and Management of Engineering Processes, Springer, London, pp.3-14, January 2010, DOI:10.1007/978-1-84996-199-8_1

[7] R. Elmasri and S. B. Navathe, Fundamentals of Database Systems, 7th ed., Pearson, Boston, 2016.

[8] J. Peckham and F. Maryanski, Semantic Data Models, ACM Computing Surveys, vol.20, no.3, pp.153–189, Sept. 1988. DOI:10.1145/62061.62062

[9] R. T. Yarlagadda and H. H. Syed, "Data models in information technology," International Journal of Innovations in Engineering Research and Technology, vol.3, no.2, Feb. 2016.

[10] M. West, "Some types and uses of data models," in Developing High Quality Data Models, pp.23-36, Elsevier Inc., 2011. DOI:10.1016/C2009-0-30508-5

[11] H. Kohen and D. Dori, "Improving conceptual modeling with object-process methodology stereotypes." Appl. Sci. 2021, 11, 2301. DOI:10.3390/ app11052301

[12] S. Al-Fedaghi, "UML modeling to TM modeling and back," International Journal of Computer Science and Network Security, vol.21, no.1, pp.84–96, 2021. DOI:10.22937/IJCSNS.2021.21.1.13

[13] C. D. Tupper, "Model constructs and model types," in Data Architecture From Zen to Reality, pp.207-221, Elsevier Inc., 2011. DOI:10.1016/B978-0-12-385126-0.00011-5

[14] Sparx Systems, "Database models," Enterprise Architect, March 6, 2017. https://sparxsystems.com/resources/user-guides/model-domains/database-models.pdf

[15] G. B. and G. Beconytė, "Entity-relationship modelling and cartographic transcription," Geodezija ir Kartografija, vol.29, no.3, pp.78-83. DOI: 10.1080/13921541.2003.10552997

[16] T. Nyerges, "Conceptual data models," ed. J. P. Wilson, The Geographic Information Science and Technology Body of Knowledge, 1st Quarter 2017. DOI: 10.22224/gistbok/2017.1.3

[17] S. Spaccapietra, C. Parent, C. Vangenot, and N. Cullot, "On using conceptual modeling for ontologies," Proc. Web Information Systems Workshops, LNCS 3307, Springer-Verslag, 2004.

[18] J. M. Osborne, "Planning, philosophy of filemaker," Blog, August 8, 2017. https://philosophyoffilemaker.com/zoom.php?recid=421

[19] M. M. Kashyap, "Likeness between Ranganathan's postulations-based approach to knowledge classification and entity relationship data modelling approach," Knowledge Organization, vol.30, no.1, pp.1-19, 2003.

[20] B. Thalheim, (2000). Entity-Relationship Modeling: Foundations of Database Technology. Berlin: Springer-Verlag.

[21] M. West, "Developing high quality data models, ver. 2.0, no. 2.1, EPISTLE (the European Process Industries STEP Technical Liaison Executive, Sep. 2003.

[22] S. Hitchman, "The entity relationship model and practical data modeling," Journal of Conceptual Modeling, issue 31, 2004.

[23] P. P. Chen, "Entity-relationship modeling: Historical events, future trends, and lessons learned," in Software Pioneers: Contributions to Software Engineering, eds. M. Broy, and E. Denert, pp. 100-114, Springer-Verlag, Berlin, 2002.

[24] H. S. Leonard and N. Goodman, "The calculus of individuals and its uses," The Journal of Symbolic Logic, vol.5, no.2, pp.45-55, June 1940.

[25] S. K. Milton and E. Kazmierczak, "An ontology of data modelling languages: A study using a common-sense realistic ontology," Journal of Database Management, vol.15, no.2, pp.19-38, 2004. DOI: 10.4018/jdm.2004040102

[26] T. J. Green, "Conceptual modeling using the entity-relationship model," lecture notes, Department of Computer Science, University of California, Davis. https://web.cs.ucdavis.edu/~green/courses/ecs165a-w11/2-er.pdf accessed Aug. 1, 2021.

[27] S. K. Karukonda, "Entity relationship approach to knowledge-based systems," LSU Historical Dissertations and Theses, 4575, 1988. https://digitalcommons.lsu.edu/gradschool_disstheses/4575

[28] I. Y. Song and P. P. Chen, "Entity relationship model," Encyclopedia of Database Systems, 2009 ed., eds L. Ling and M. Tamer Özsu. DOI:10.1007/978-0-387-39940-9_148

[29] B. C. van Fraassen, An Introduction to the Philosophy of Time and Space, Nousoul Digital Publishers, 2015. https://www.princeton.edu/~fraassen/BvF%20-%20IPTS.pdf

[30] S. J. Mason, "Feedback theory: Further properties of signal flow graphs," Proc. IRE, vol.44, no.7, pp.920-926, 1956.

[31] T.R. Gruber, "Toward principles in the design of ontologies used for knowledge sharing," Technical Report 93-04, Knowledge Systems Laboratory, Palo Alto, CA, Stanford University.

[32] D. P. Dhabe, M. S. Patwardhan, S. P. Pundlik, M. Dhore, B. V. Barbadekar and H. K. Abhyankar, "Articulated entity relationship (AER) diagram for complete automation of relational database normalization," International Journal of Database Management Systems, vol.2, no.2, pp.84-100, May 2010. DOI: 10.5121/ijdms.2010.2206

[33] S. R. Ruppel, System Behavior Models: A Survey of Approaches, Thesis, Naval Postgraduate School, Monterey, CA, June 2016. https://apps.dtic.mil/sti/pdfs/AD1026811.pdf